\documentstyle[twoside,fleqn,espcrc2,epsf]{article}

\newcommand{\AmS}{{\protect\the\textfont2
  A\kern-.1667em\lower.5ex\hbox{M}\kern-.125emS}}

\title{
The equation of state for two flavor QCD at finite density
\thanks{Presented by S. Ejiri. This work is supported by 
BMBF under grant No.06BI102, DFG under grant FOR 339/2-1 
and PPARC grant PPA/a/s/1999/00026.
}}

\author{ S. Ejiri\rlap,
\address{Fakult\"{a}t f\"{u}r Physik, Universit\"{a}t 
Bielefeld, D-33615 Bielefeld, Germany}
C.R. Allton\rlap,\address{Department of Physics, University of 
Wales Swansea, Singleton Park, Swansea, SA2 8PP, U.K.} 
S.J. Hands\rlap,$^{\rm b}$ 
O. Kaczmarek\rlap,$^{\rm a}$  F. Karsch\rlap,$^{\rm a}$ 
E. Laermann\rlap,$^{\rm a}$ and C. Schmidt$^{\rm a}$}

\begin{document}

\begin{abstract}
We discuss the equation of state for QCD at 
non-zero temperature and density. We present results of a simulation 
for QCD with 2 flavors of p4-improved staggered fermions. 
Derivatives of $\ln {\cal Z}$ with respect to 
quark chemical potential $\mu_q$ up to fourth order are calculated, 
enabling estimates of the pressure, quark number density and 
associated susceptibilities as functions of $\mu_q$ via a Taylor 
series expansion. We also discuss the radius of convergence of 
the expansion as a function of temperature. 
It is found that the fluctuations in the quark number density 
increase in the vicinity of the phase transition temperature and 
the susceptibilities start to develop a pronounced peak 
as $\mu_q$ is increased. 
This suggests the presence of a critical endpoint in the 
$(T, \mu_q)$ plane.
\vspace{1pc}
\end{abstract}

\maketitle

\section{Introduction}
\label{sec:intro}

In the last few years, remarkable progress has been made 
in the theoretical study of QCD thermodynamics with small but 
non-zero quark chemical potential $\mu_q$. In particular, numerical 
studies of the phase structure in the $(T, \mu_q)$ plane \cite{Fod,us02,dFP} 
provide important information for heavy-ion collision experiments. 
One of the most interesting investigations is finding 
the critical endpoint, at which the crossover phase transition 
line becomes first order. It might be 
possible to detect the endpoint experimentally via event-by-event 
fluctuations, since 
the number density fluctuation should be large around the endpoint
\cite{RSS}. 
The fluctuation should be in proportion to the susceptibility of 
the quark number density, hence the study of the equation state at 
non-zero baryon number density is very important \cite{AHMJK,TK}. 

The quark number susceptibility is given by the second derivative 
of pressure with respect to $\mu_q$. 
Several studies of the quark number susceptibility 
have been performed at $\mu_q=0$ \cite{Gott}, 
and measurements of pressure and energy density at 
$\mu_q \neq 0$ were done by \cite{FKS} using the reweighting method, 
which allow us to investigate thermodynamic properties at 
$\mu_q \neq 0$. This approach, however, does not work 
for large $\mu_q$ and large lattice size due to the sign problem. 
Here, we adopt the following strategy. 
We compute the derivatives of thermodynamic quantities such as 
pressure, quark number susceptibility etc., with respect to $\mu_q$ 
at $\mu_q=0$, and determine the Taylor expansion coefficients in $\mu_q$ 
\cite{us02,GG}, in which the sign problem does not arise. 

We discussed last year the relation between the line of 
constant pressure (energy density) and the phase transition line 
calculating the second derivative of pressure \cite{us02}. 
In this study \cite{us03}, we proceed to fourth order derivatives. 
Then, we can for the first time investigate the $\mu_q$-dependence 
of the quark number susceptibility near $\mu_q=0$. 
(Because the pressure is an even function of $\mu_q$, 
the $\mu_q^2$-term is leading and $\mu_q^4$-term is the next to leading,
and in fact only these two terms are non-zero 
in the high temperature limit.)
By estimating the change of the susceptibility, 
we discuss the possibility of the existence of the critical 
end point in the phase diagram of $T$ and $\mu_q$.

\begin{figure}[t]
\centerline{
\epsfxsize=6.4cm\epsfbox{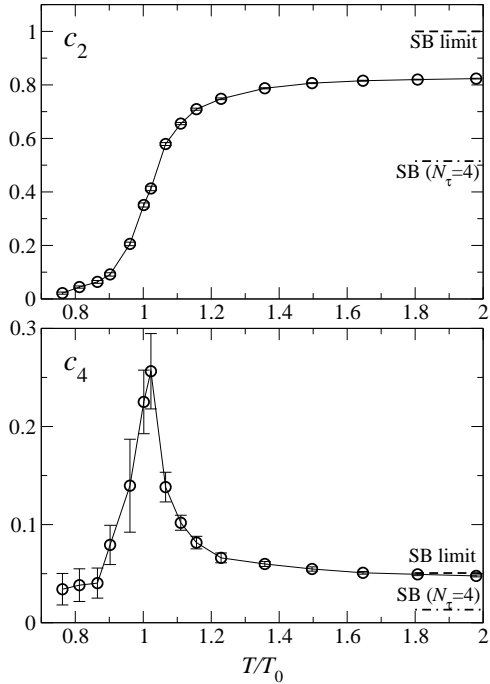}
}
\vspace*{-11mm}
\caption{
Coefficients of Taylor expansion, $c_2$ (upper) and $c_4$ (lower).
$T_0$ is $T_c$ at $\mu_q=0$.
}
\vspace*{-4mm}
\label{fig:c2c4}
\end{figure}

\section{Taylor expansion in $\mu_q$}
\label{sec:expansion}

Pressure is given in terms of the grand partition function 
${\cal Z}(T, V, \mu_q)$ by
$p/T^4=(1/VT^3)\ln{\cal Z}.$
However, the direct calculation of $\ln {\cal Z}$ is difficult, hence 
most of the work done at $\mu_q=0$ for the calculation of pressure 
is done by using the integral method \cite{KLP,CPPACS}, 
where the first derivative of 
pressure is computed by simulations, and the pressure is obtained by 
integration along a suitable integral path. 
For $\mu_q \neq 0$, direct Monte Carlo simulation is not applicable; 
in this case we proceed by computing higher order derivatives of 
pressure with respect to $\mu_q/T$ at $\mu_q=0$, and then 
estimate $p(\mu_q)$ using a Taylor expansion, 
\begin{eqnarray}
\left. \frac{p}{T^4} \right|_{T,\mu_q} 
= \left. \frac{p}{T^4} \right|_{T,0} 
+ \sum_{n=1}^{\infty} c_n(T) \left( \frac{\mu_q}{T} \right)^n ,
\label{eq:taylorcont}
\end{eqnarray}
where $c_n=(1/n!) \partial^n(p/T^4)/\partial(\mu_q/T)^n |_{\mu_q=0}$.
These derivatives can be computed by the random noise method, which 
saves CPU time, and also it can be proved that 
the odd terms are exactly zero. 
Furthermore, we do not need simulations at $(T, \mu_q)=(0,0)$ 
for subtraction term of $p$ to normalize the value of $p$, since 
the derivatives of $p|_{T=0,\mu_q=0}$ with respect to $\mu_q$ are, 
of course, zero. This also reduces CPU time.

The quark number density $n_q$ and quark number susceptibility 
$\chi_q$ are obtained by the derivatives of pressure;
\begin{eqnarray}
\frac{n_q}{T^3} = \frac{\partial (p/T^4)}{\partial (\mu_q/T)}, 
\hspace{5mm}
\frac{\chi_q}{T^2}
= \frac{\partial^2 (p/T^4)}{\partial (\mu_q/T)^2}
\label{eq:chiq}.
\end{eqnarray}
We compute the pressure up to $\mu_q^4$-term using 2 flavors of 
p4-improved staggered fermions \cite{HKS} at $ma=0.1$ 
on a $16^3 \times 4$ lattice. 
Then the quark number density and quark number susceptibility are 
obtained up to $O(\mu_q^3)$ and $O(\mu_q^2)$, respectively. 
The details of the simulations are given in \cite{us03}. 
In Fig.~\ref{fig:c2c4}, we plot the data for $c_2$ (upper) and 
$c_4$ (lower). Both of them are very small at low temperature and 
approach the Stefan-Boltzmann (SB) limit in the high temperature 
limit, as we expected. 
The remarkable point is a strong peak of $c_4$ around $T_c$. 

We can understand this peak by the following two issues. 
One is a prediction from the hadron resonance gas model \cite{KRT}, 
which is an effective model of the free hadron gas in the low temperature 
phase. 
The model study predicts $c_4/c_2=0.75$ and our results are consistent 
with this prediction for $T < T_c$; in fact, as $T$ increases $c_4/c_2$ 
remains constant until $T \approx T_c$, 
whereupon it approaches the SB limit. 

The other point is from a discussion of the convergence radius of 
the Taylor expansion. 
We expect that the crossover transition at $\mu_q=0$ changes to 
first order transition at a point $\mu_q/T_c \sim O(1)$ \cite{Fod,FK03}. 
Then, the analysis by the Taylor expansion 
must break down in that regime, i.e. the convergence radius may be 
smaller than the critical endpoint.
We define estimates for the convergence radius by 
$\rho_n \equiv \sqrt{|c_n / c_{n+2}|}$.
We compute $\rho_0$ and $\rho_2$ from $p/T^4(0)$, $c_2$ and $c_4$. 
It is found that both $\rho_0$ and $\rho_2$ are quite large at 
high temperature as expected from SB limit,
$\rho_2^{SB} \simeq 2.01, \rho_4^{SB} \simeq 4.44.$ On the other hand, 
around $T_c$, these are $O(1)$, since $c_2$ and $c_4$ are of the same order, 
so that our results around $T_c$ suggest a singular point in the 
neighborhood of $\mu_q/T_c =1$.

\begin{figure}[t]
\centerline{
\epsfxsize=6.1cm\epsfbox{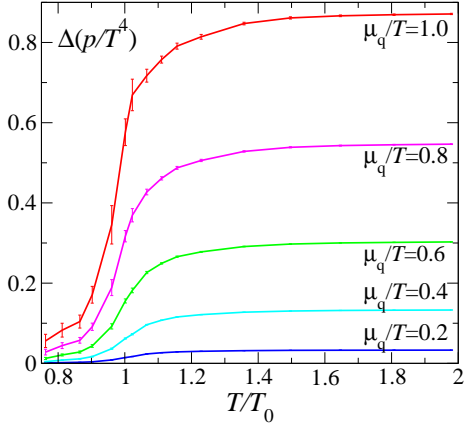}
}
\vspace*{-11mm}
\caption{
Difference of pressure from $\mu_q=0$ as a function of $T$ 
for each fixed $\mu_q/T$.
}
\vspace*{-4mm}
\label{fig:pres}
\end{figure}

\begin{figure}[t]
\centerline{
\epsfxsize=6.0cm\epsfbox{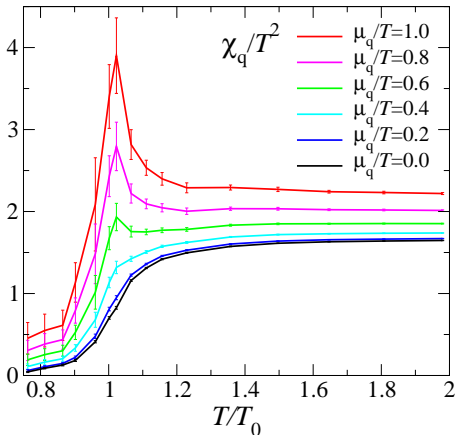}
}
\vspace*{-11mm}
\caption{
Quark number susceptibility as a function of $T$ for each fixed $\mu_q/T$.
}
\vspace*{-4mm}
\label{fig:nsus}
\end{figure}

\section{Equation of state at $\mu_q \neq 0$} 
\label{sec:eos}
Next, we calculate pressure and quark number susceptibility in a range 
of $0 \leq \mu_q/T \leq 1$ which is within the radius of convergence 
discussed above, using the data of $c_2$ and $c_4$; 
$\Delta(p/T^4)=p(T,\mu_q)/T^4-p(T,0)/T^4=
c_2(\mu_q/T)^2+c_4(\mu_q/T)^4+O(\mu_q^6)$, 
and $\chi_q/T^2=2c_2+12c_4(\mu_q/T)^2+O(\mu_q^4)$. 

We draw $\Delta (p/T^4)$ for each fixed $\mu_q/T$ in Fig.~\ref{fig:pres} 
and find that the difference from $p|_{\mu_q=0}$ is very small 
in the interesting regime for heavy-ion collisions, 
$\mu_q/T \approx 0.1$ (RHIC) and $\mu_q/T \approx 0.5$ (SPS), 
in comparison with the value at $\mu_q=0$, e.g. SB value for 2 flavor QCD 
at $\mu_q=0$: $p^{SB}/T^4 \simeq 4.06$. The finite density effect of 
pressure at $\mu_q/T=0.1$ is only $1\%$. 
This conclusion is consistent with that inferred from just the $O(\mu_q^2)$ 
term \cite{us02}. Also, the result is qualitatively consistent with that 
of \cite{FKS} obtained by the reweighting method. 

Figure \ref{fig:nsus} is the result of $\chi_q/T^2$ for fixed $\mu_q/T$. 
We find a pronounced peak for $m_q/T > 0.5$, whereas $\chi_q$ does 
not have a peak for $\mu_q=0$.
This suggests the presence of a critical endpoint in the $(T,\mu_q)$ plane.

This discussion can be easily extended to the charge fluctuation 
$\chi_C$.
The spike of $\chi_C$ at $T_c$ is weaker than that of 
$\chi_q$, which means the increase of the charge fluctuation is smaller 
than that of the number fluctuation as $\mu_q$ increases \cite{us03}.


\begin{thebibliography}{99}

\bibitem{Fod} Z. Fodor and S.D. Katz, JHEP 0203 (2002) 014. 

\bibitem{us02}
C.R. Allton {\it et.al.}, Phys. Rev. {\bf D66} (2002) 074507.

\bibitem{dFP}
P. de Forcrand and O. Philipsen, Nucl. Phys. {\bf B642} (2002) 290; 
{\tt hep-lat/0307020};
M. D'Elia and M.-P. Lombardo, Phys. Rev. {\bf D67} (2003) 014505.

\bibitem{RSS}
M. Stephanov, K. Rajagopal and E. Shuryak,
Phys. Rev. Lett. {\bf 81} (1998) 4816.

\bibitem{AHMJK}
M. Asakawa, U. Heinz and B. M\"uller,
Phys. Rev. Lett. {\bf 85} (2000) 2072;
S. Jeon and V. Koch, 
Phys. Rev. Lett.  {\bf 85} (2000) 2076;
{\tt hep-ph/0304012}.

\bibitem{TK}
T. Kunihiro, {\tt hep-ph/0007173}.

\bibitem{Gott} S. Gottlieb {\it et al.}, 
Phys.\ Rev.\ {\bf D38}\ (1988)\ 2888; 
C. Bernard {\it et.al.}, 
Nucl. Phys. B (Proc. Suppl.) {\bf 119} (2003) 523.

\bibitem{GG}
R.V. Gavai and S. Gupta, Phys. Rev. {\bf D68} (2003) 034506.

\bibitem{FKS}
Z. Fodor, S.D. Katz and K.K. Szab\'o, {\tt hep-lat/0208078}.

\bibitem{us03}
C.R. Allton {\it et.al.}, Phys. Rev. {\bf D68} (2003) 014507.

\bibitem{KLP}
F. Karsch, E. Laermann and A. Peikert, Phys. Lett. {\bf B478} (2000) 447.

\bibitem{CPPACS} 
CP-PACS,~Phys.~Rev.~{\bf D64}~(2001)~074510.

\bibitem{HKS} U.M. Heller, F. Karsch, and B. Sturm, 
Phys. Rev. D60 (1999) 114502.

\bibitem{KRT}
F. Karsch, K. Redlich and A. Tawfik, 
{\tt hep-ph/0303108}; {\tt hep-ph/0306208}.

\bibitem{FK03} 
C. Schmidt {\it et.al}, Nucl. Phys. B (Proc. Suppl.) {\bf 119} (2003) 517;
F. Karsch {\it et.al}, these proceedings.



\end{thebibliography}
\end{document}